%% file: pap2.tex
\documentclass[seceqn]{elsart}
\usepackage{macros}
\usepackage{macros_pap2}
\usepackage[dvips]{epsfig}
\begin{document}
\input title.tex
\input sect1.tex
\input sect2.tex
\input sect3.tex
\input sect4.tex
\input sect5.tex

\input sect6.tex
\input sect7.tex
\begin{appendix}
\input appa.tex
\end{appendix}
\bibliography{pap2}      % or whatever your .bib file is
\bibliographystyle{h-elsevier}    % if you use h-elsevier.bst
\end{document}

%% file: title.tex
\begin{frontmatter}

\begin{flushright}
DESY 01-111 \\
SHEP 01/22 \\
hep-lat/0108018
\end{flushright}

\title{
Heavy Quark Effective Theory at one-loop order: An explicit example
}
\vbox{
\centerline{
\epsfxsize=2.5 true cm
\epsfbox{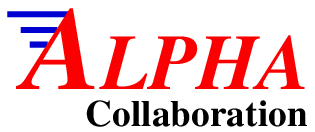}}
}
\author[soton]{Martin Kurth}
\author[DESY]{Rainer Sommer}
\address[soton]{
Department of Physics and Astronomy, University of Southampton \\
Southampton SO17 1BJ, UK
}
\address[DESY]{
Deutsches Elektronen-Synchrotron, DESY \\
Platanenallee 6, D-15738 Zeuthen, Germany
}

\maketitle
\begin{abstract}

We consider correlation functions containing the axial current of one light
and one heavy quark in the static approximation as well as in full QCD,
using the lattice regularization. Up to one-loop order of perturbation theory,
we study the difference between the full and the effective theory in the
continuum limit. In the full theory we find a  term non-analytic in $1/m$,
revealing the asymptotic character of the $1/m$-expansion. In general, deviations
from the $m\to\infty$ limit turn out to be small and are well described by
the first non-trivial terms when $m$ is a factor 2-3 above the external scale.
We also investigate the mass dependence of discretization errors, and find
that the behaviour of the correlation functions at finite lattice spacing
differs significantly from that in the continuum limit when the quark mass
is large.

\end{abstract}

\keyword{
heavy quark effective theory; matching; 
lattice QCD}\\
\PACS{11.15.Ha; 12.38.Bx; 12.38.Gc; 12.39.Hg}
\endkeyword

\end{frontmatter}
\cleardoublepage

%%% Local Variables: 
%%% mode: latex
%%% TeX-master: "pap2"
%%% End: 

%% file: sect1.tex
\section{Introduction \label{s_Introduction}}

Heavy quark effective theory (HQET) is an important tool for heavy quark
phenomenology. As a means of decoupling the light degrees of freedom from
the mass and spin of heavy quarks (with masses that are much larger than
$\Lambda_{\rm QCD}$), it is a powerful method to calculate e.g. the
heavy hadron spectrum and weak decay form factors. For a review of
HQET see~\cite{Neubert:1994mb}. An interesting object
to study is the axial current of one light and one heavy quark, as the
B-meson decay constant $\fb$ is defined by a matrix element of such
a current.

QCD matrix elements have to be evaluated non-perturbatively, which means
that the lattice regularization has to be used. As only finite volumes can
be treated numerically, and the volume must be large enough to accommodate
the physical system such that finite size effects are negligible, current
com\-puting power enforces a lower bound on accessible lattice spacings, i.e.
an upper bound on the cutoff. As, on today's computers, this cutoff cannot
be chosen much larger than the $b$ quark mass, the simulation of bottom
flavoured particles is not possible in full QCD, and the use of effective
theories like HQET becomes a necessity.

In a previous paper~\cite{Kurth:2000ki}, we presented 
the outline of a method for
the non-perturbative renormalization of the heavy-light axial current 
in the static approximation,
which corresponds to the leading order in the heavy quark expansion,
where we used Schr{\"o}dinger functional techniques. 
In this framework, one considers QCD in a finite volume of linear size $L$
and suitable boundary conditions. The length scale $L$ takes over the r\^ole
played by external momenta $p_\mu$ in transition amplitudes in infinite 
volume.
HQET emerges in the large mass limit at fixed $L$. 
When no other scale is present, as it is the case in perturbation theory,
one has to consider the $z\to\infty$ limit with $z=m\,L$ and some definition
of the heavy quark mass, $m$. The predictive power of HQET comes from the
fact that the matching of QCD and HQET does not depend on the external 
states (momenta). It is therefore also independent of the boundary 
conditions imposed  (and $L$). For this reason we may use the 
Schr{\"o}dinger functional to study the relation between the static 
approximation and full QCD. The advantage is that on the one hand
gauge invariant, infrared finite correlation functions may be computed 
in perturbation theory (rather easily) and on the other hand 
its  lattice-regulated version is very well suited for non-perturbative 
Monte Carlo simulations \cite{impr:lett}. 

In this paper, we investigate the difference between the static approximation
and full QCD in the {\em continuum} limit,
for the specific correlation functions defined 
in~\cite{Kurth:2000ki}, both at tree level and at one-loop order of
perturbation\linebreak
theory. For the first time (to our knowledge),
it is explicitly demonstrated that gauge invariant, infrared finite 
correlation functions are described by HQET in the limit $m\to \infty$.
Indeed, we observe that the $1/m$-expansion is only an asymptotic expansion;
small non-analytic terms $\exp(-z)$ are present in the full theory.

A second point we look at in this paper is the
mass dependence of discretization errors, after applying Symanzik's
improvement programme~\cite{impr:Sym1} to actions, boundary fields, and the
axial current.

We use the notation of reference~\cite{Kurth:2000ki}. This paper is organized 
as follows. In section 2, the basic definitions
of the static quark action, both in the continuum and on the lattice, are
given. Schr{\"o}dinger functional boundary conditions and the correlation
functions we use are also explained in that section. In section 3, the
renormalization schemes we use in the static approximation and in full
QCD are introduced. The investigation of finite mass effects, i.e. the
difference between the full and effective theories, is carried out in
section 4. Section 5 is devoted to the study of discretization errors and
their dependence on the heavy quark mass. Our results are summarized in
section 6. An appendix contains the relevant one-loop Feynman diagrams.

%%% Local Variables: 
%%% mode: latex
%%% TeX-master: pap2
%%% End: 

%% file: sect2.tex
\section{Correlation functions \label{s_Correlation}}

\subsection{Static quarks \label{s_Static}}

Static quarks can be treated as two-component objects. To simplify the 
notation, they will be described as four-component spinors here,
satisfying the constraint
\be
  \label{e_constraint}
  P_{+}\heavy=\heavy\,,\quad \heavyb P_{+}=\heavyb\,,\quad 
  P_{+}=\frac12(1+\gamma_0)\,.
\ee
The static quark action is
\be
  \label{e_ContLag}
  \Sstat=\int{\rm d}^4x\,\heavyb(x){\rm D}_0\heavy(x).
\ee
Following \cite{stat:eichhill1}, we discretize the static quark action as
\be
  \label{e_LatLag}
  \Sstat=a^4\sum_x \heavyb(x)\nabstar{0} \heavy(x)\,.
\ee

\subsection{The Schr\"odinger functional
\label{ss_schrodinger}}

The Schr\"odinger functional consists of a finite space-time volume
$T\times L\times L\times L$,
with specifically chosen boundary conditions. Explicitly, the gauge
field is chosen to be periodic in space and to satisfy Dirichlet
boundary conditions in time,
\be
  A_k(x)|_{x_0=0}=C(\vecx),\qquad
  A_k(x)|_{x_0=T}=[\projector C'](\vecx)\,,
\ee
where $C(\vecx)$ and $C'(\vecx)$ are fixed boundary fields,
and $\projector$ projects onto the gauge invariant content of $C'(\vecx)$
(see \cite{SF:LNWW}). Here we choose $C=C'=0$,
leading to the boundary conditions
\be
  U(x,k)|_{x_0=0}=1\,,\quad
  U(x,k)|_{x_0=T}=1
\ee
for the lattice gauge field.

The light quark field is chosen to be periodic up to a phase in the
three space directions,
\be
  \label{e_lambdabound}
  \psi(x+L\hat{k})=\rme^{i\theta}\psi(x),
  \qquad
  \psibar(x+L\hat{k})=\psibar(x)\rme^{-i\theta},
\ee
where $\theta$ is kept as a free parameter.

As for the gauge field, Dirichlet boundary conditions are imposed on
the light quark field,
\be
  P_{+}\light(x)|_{x_0=0}=\rhol(\vecx),\quad
  P_{-}\light(x)|_{x_0=T}=\rholprime(\vecx),
\ee
and
\be
  \lightb(x)P_{-}|_{x_0=0}=\rholb(\vecx),\quad
  \lightb(x)P_{+}|_{x_0=T}=\rholbprime(\vecx),
\ee
as well as on the static quark field,
\be
  \heavy(x)|_{x_0=0}=\rhoh(\vecx),\quad
  \heavyb(x)|_{x_0=T}=\rhohbprime(\vecx).
\ee
Note that no projectors are necessary in the heavy quark case because
of \eq{e_constraint}. For the same reason, we have
$P_{-}\heavy(x)=0$. Spatial boundary conditions do not need to be
discussed for the static quarks
since they do not propagate in space. The quark field boundary
conditions given above can be applied both in the continuum and on the 
lattice.

The Schr\"odinger functional action, including boundary and $\Oa$
improvement terms, can be found in~\cite{SF:LNWW,SF:stefan1,Kurth:2000ki}.
In this paper, it is
understood that all quantities are calculated using the $\Oa$ improved
action.

\subsection{Correlation functions \label{ss_Correlation}}

\begin{figure}
  \noindent
  \begin{center}
  \begin{minipage}[b]{.3\linewidth}
    \centering\epsfig{figure=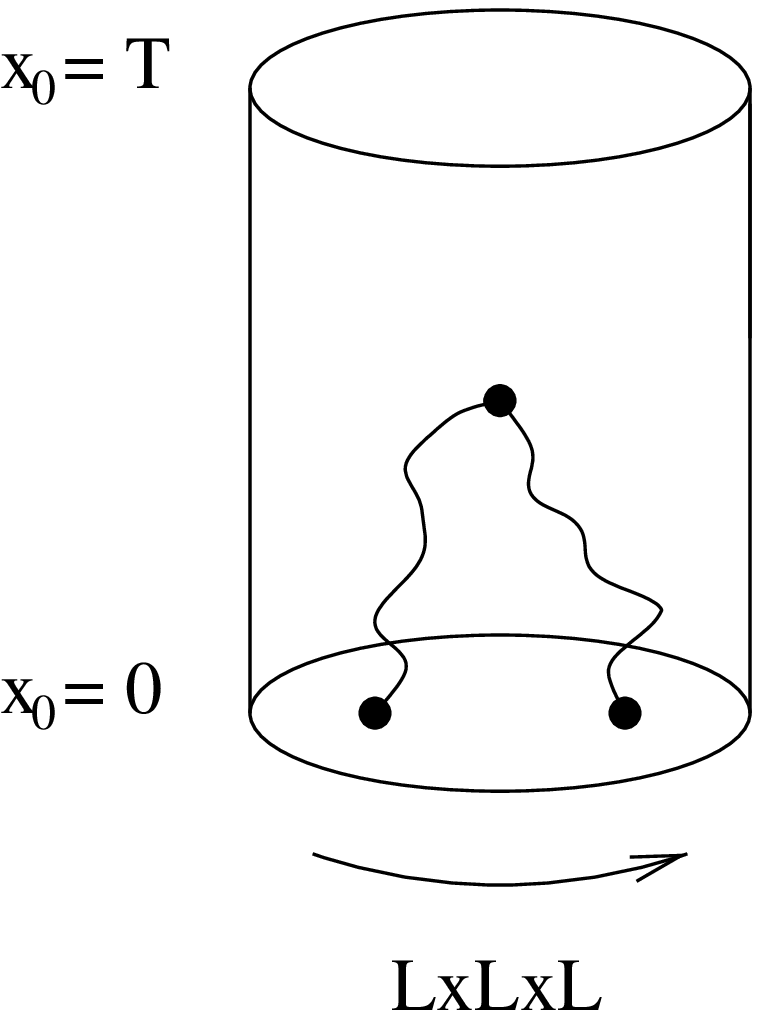,width=\linewidth}
  \end{minipage}
  \hspace{20mm}
  \begin{minipage}[b]{.3\linewidth}
    \centering\epsfig{figure=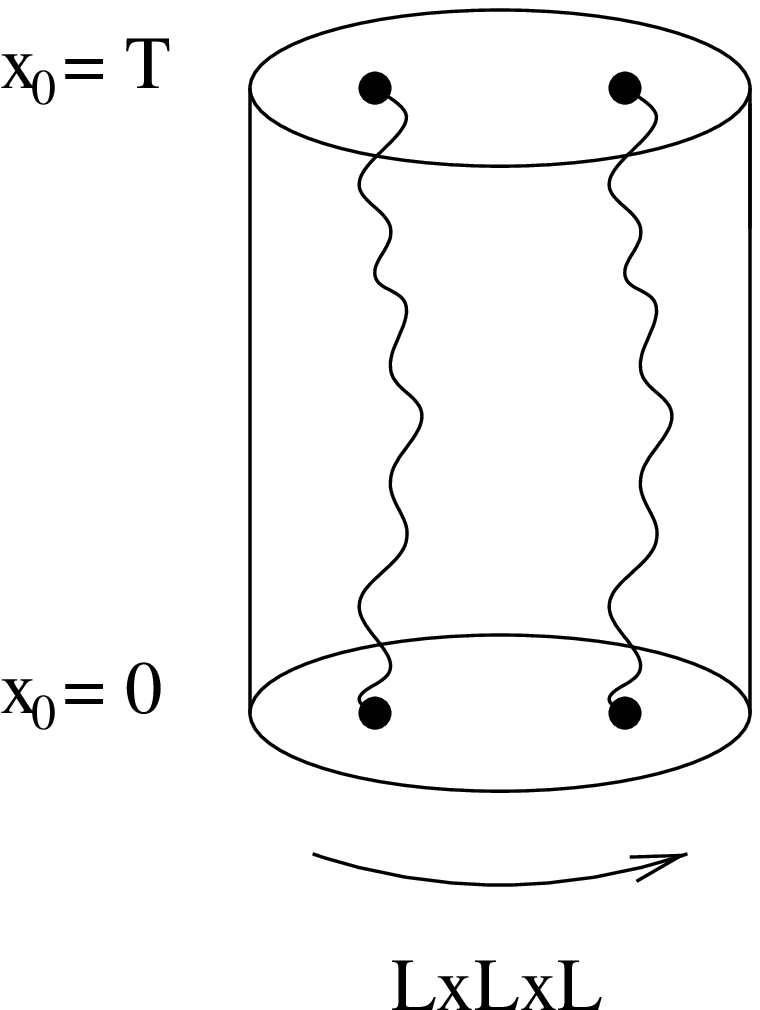,width=\linewidth}
  \end{minipage}
  \end{center}
  \caption[The correlation functions $\fa$ and
    $\fone$]{\footnotesize Schematic drawing of the correlation functions
    $\fa$ (left) and $\fone$ (right). The dot in the middle of the
    left diagram symbolizes the axial current at $x_0$, the other dots 
    represent the boundary quark fields. \vspace{0.6cm}
    }
  \label{f_CylFaF1}
\end{figure}

In the following, we study two different correlation
functions. The function $\fa$ contains the time
component of a static-light axial current
\be
  A_0^{\rm stat}(x)=\overline{\psi}_1(x)\gamma_0\gamma_5\heavy(x)
\ee
or a light-light axial current
\be
  A_0(x)=\overline{\psi}_1(x)\gamma_0\gamma_5\psi_2(x),
\ee
where the indices $1$ and $2$ label two flavours of relativistic quarks.
It is correlated with a suitable combination of ``boundary quark fields''
to yield the diagram illustrated  in the left  
of \Fig{f_CylFaF1}
upon Wick-contraction. In addition a boundary-to-boundary correlation
$\fone$, shown on the right of the figure is needed. 

With the boundary quark fields $\zeta,\overline{\zeta}$ 
(see  \cite{Kurth:2000ki} for their proper definition), 
the correlation functions
\be
  \fastat(x_0)=-{1\over 2}\int\rmd^3\vecy\,\rmd^3\vecz\,
  \langle\Astat(x)\zetahb(\vecy)\gamma_5\zeta_1(\vecz)\rangle
\ee
and
\be
  \fonestat=-{1\over{2L^6}}\int
  \rmd^3\vecu\,\rmd^3\vecv\,\rmd^3\vecy\,\rmd^3\vecz\,
  \langle\overline{\zeta}_1\kern1pt'(\vecu)\gamma_5\zetahprime(\vecv)
  \zetahb(\vecy)\gamma_5\zeta_1(\vecz)\rangle\,
\ee
can be introduced as in~\cite{Kurth:2000ki}. In addition, the corresponding 
functions 
in full QCD,
\be
  \fa(x_0)=-{1\over 2}\int\rmd^3\vecy\,\rmd^3\vecz\,
  \left\langle A_0(x)\zetalbtwo(\vecy)\gamma_5\zetalone(\vecz)\right\rangle,
\ee
and
\be
  \fone=-{1\over{2L^6}}\int
  \rmd^3\vecu\,\rmd^3\vecv\,\rmd^3\vecy\,\rmd^3\vecz\,
  \left\langle\zetalbprimeone(\vecu)\gamma_5\zetalprimetwo(\vecv)
  \zetalbtwo(\vecy)\gamma_5\zetalone(\vecz)\right\rangle
\ee
are used. They are discretized as in references~\cite{impr:pap4,impr:pap5},
including $\Oa$ improvement terms
\be
  a\castat\delta\Astat=a\castat \overline{\psi}_1
       \frac12(\lnab{j}+\lnabstar{j})\gamma_j\gamma_5\heavy
\ee
and
\be
  a\ca\delta A_0=a\ca\frac12(\partial_0^{\ast}+\partial_0)
    \left\{ \overline{\psi}_1\gamma_5\psi_2 \right\},
\ee
with the improvement coefficients
\be
  \castat=g_0^2c_{\rm A}^{\rm stat(1)}+{\rm O}(g_0^4),
  \qquad c_{\rm A}^{\rm stat(1)}=-{1\over{4\pi}}\times 1.00(1)
\ee
and
\be
  \ca=g_0^2c_{\rm A}^{(1)}+{\rm O}(g_0^4),\qquad 
  c_{\rm A}^{(1)}=-0.00567(1)C_{\rm F},
\ee
where the coefficient $c_{\rm A}^{\rm stat(1)}$ has been calculated
in~\cite{MORNINGSTAR1} using NRQCD 
(a calculation using a different method can
be found in~\cite{Ishikawa}), and the coefficient $c_{\rm A}^{(1)}$ has been
taken from~\cite{impr:pap2}.

With these correlation functions, the ratios
\be
  X_{\rm I}(g_0,L/a)={{\fastatimpr(T/2)}\over{\sqrt{\fonestat}}}
\ee
and
\be
  Y_{\rm I}(g_0,z,L/a)={{\faimpr(T/2)}\over\sqrt{\fone}}
\ee
are defined, where the index $\rm I$ means that the $\Oa$ improved
axial currents are used, and $z$  parametrizes the mass dependence as
defined in subsection~\ref{ss_RenormSchemes} below.
In the ratios $X_{\rm I}$ and $Y_{\rm I}$,
renormalization constants for the light and static quark boundary
fields cancel as well as a $1/a$ divergence due to a heavy quark
mass counterterm in the static
approximation (see~\cite{Kurth:2000ki} for details).

For our perturbative analysis, we expand the ratios to one-loop order,
\be
  X_{\rm I}(g_0,L/a)=X_{\rm I}^{(0)}(L/a)
    +X_{\rm I}^{(1)}(L/a)g_0^2+{\rm O}(g_0^4),
\ee
\be
  Y_{\rm I}(g_0,z,L/a)=Y_{\rm I}^{(0)}(z,L/a)
   +Y_{\rm I}^{(1)}(z,L/a)g_0^2+{\rm O}(g_0^4).
\ee

%%% Local Variables: 
%%% mode: latex
%%% TeX-master: "pap2"
%%% End: 

%% file: sect3.tex
\section{Renormalization and matching \label{s_RenMatch}}

When taking the continuum limit, renormalization of quark masses,
couplings, and composite operators is required. Bare lattice quark
masses $m_0$
need an additive renormalization
\be
  \mq=m_0-\mcrit
\ee
due to the chiral symmetry breaking term in the Wilson fermion action, 
de\-fining a reduced quark mass $\mq$. The critical mass $\mcrit$ is
determined by the requirement that in the case of two mass-degenerate
quarks the PCAC mass is zero at $\mq=0$. 
In perturbation theory, the
critical mass can be expanded as 
\be
  \mcrit=g_0^2\mcritone+{\rm O}(g_0^4).
\ee
Depending on the precise definition of the PCAC mass,
$\mcrit$ (and therefore also $\mcritone$) depend on the 
lattice spacing. However, in perturbation theory
one may take the limit where external momentum scales are small compared
to the cutoff $1/a$, and extrapolate  
the coefficient $\mcritone$ to its continuum limit.
This gives~\cite{impr:pap2}
\be
  a \mcritone=-0.2025565(1)\times\cf,
\ee
a value which will be used also in \sect{s_Cutoff} at finite 
values of $a$.
Throughout this paper, we use $\mq=0$ for the light quark field $\psi_1$.

\subsection{Renormalization schemes \label{ss_RenormSchemes}}

Unlike the case of full QCD, which is known to be renormalizable at all
orders of lattice perturbation theory~\cite{Reisz:1989kk}, renormalizability of
the static theory has not yet been proven. However, a large number of
calculations has been performed in the static approximation, and none of
these has given a reason to doubt renormalizability. Throughout this
paper, we assume that the static theory is renormalizable ``as usual''
by adding local counterterms to action and composite fields. 
 
At one-loop level, the ratio $X$ is divergent in the continuum limit, i.e. 
a scale dependent renormalization is required. The one-loop coefficient 
$X^{(1)}$ is
expected to diverge logarithmically, and a renormalization scheme, the
\emph{lattice MS scheme} can be defined by subtracting this logarithm,
\be
  X_{\rm lat}(\mu)=\zastatLat(\mu) X_{\rm I}(a/L),
\ee
\be
  \zastatLat(\mu)=1-\gamma_0\ln(a\mu)\gLat^2,
\ee
where
\be
  \gLat^2=g_0^2+{\rm O}(g_0^4),
\ee
and $\mu$ is the renormalization scale.
The one-loop anomalous dimension of the static axial current
\be
  \gamma_0=-{1\over{4\pi^2}}
\ee
does not depend on the renormalization scheme and
is known from~\cite{Shifman:1987sm,Politzer:1988wp}.

The ratio $Y_{\rm I}(z,a/L)$ remains finite in the continuum 
limit as long as the\linebreak
coupling and the heavy quark mass are made finite by renormalization. The 
renormalization condition used here is that, in the continuum limit, 
they equal the $\msbar$ coupling and quark masses respectively, at the
renormalization scale $1/L$. At one-loop
level, this means that the renormalized quantities
\be
  g_{\msbar}^2=g_0^2+{\rm O}(g_0^4)
\ee
and
\be
  \overline{m}_{2,\msbar}=\zmMSbar(1+ab_{\rm m}m_{\rm q,2})m_{\rm q,2}
\ee
are used, where $m_{\rm q,2}$ is the reduced quark mass of the second quark
flavour (remember that the mass of the first flavour is chosen to be zero),
\be
  \zmMSbar=1-\left({1\over{2\pi^2}}\ln(a/L)-0.122282\cf\right)
  g_{\msbar}^2+{\rm O}(g_{\msbar}^4)
\ee
is known from~\cite{pert:gabrielli}, and the improvement coefficient
\be
  b_{\rm m}=-{1\over{2}}-0.07217(2)C_{\rm F}g_0^2
  +{\rm O}(g_0^4)
\ee
has been calculated in~\cite{impr:pap5}.

Taking the continuum limit means to send the lattice spacing to zero while
keeping the physical box size fixed. One must thus take the limit 
${a\over L}\longrightarrow 0$, while $g_{\msbar}$ and
$ z=L\overline{m}_{2,\msbar}$
are kept at fixed values.

A finite renormalization of the light-light axial current is possible,
and we define a renormalized current $\ArenCA$ by requiring that it satisfies
the current algebra relations in the case $z=0$
(see~\cite{Boch,impr:pap4}). This renormalization
condition, together with the $\msbar$ definition of the coupling and
the quark mass, gives a renormalized ratio 
\be
  Y_{\rm CA}(z,a/L)=\za(1+\frac12 ab_{\rm A}m_{\rm q,2}) 
    Y_{\rm I},
\ee
with the perturbative renormalization constant~\cite{pert:gabrielli}
\be
  \za=1-\za^{(1)}\cf g_0^2+{\rm O}(g_0^4),\qquad
  \za^{(1)}=-0.0873435,
\ee
and the improvement coefficient~\cite{impr:pap5}
\be
  b_{\rm A}=1+0.11414(4)C_{\rm F}g_0^2+{\rm O}(g_0^4).
\ee
In principle, $Y_{\rm CA}$ should be considered also as a
function of $ g^2_\msbar$. Since at 1-loop order the dependence
on the coupling is rather trivial, we have not indicated it
explicitly.

\subsection{Matching of full and effective theories
\label{ss_Matching}}

By construction, the effective theory is expected to describe the physics
of full QCD in the limit $m\rightarrow\infty$, when this limit is taken
appropriately~\cite{Mannel:1992mc}. Beyond the classical level
this physics boundary condition translates into a \emph{matching condition}
for the two theories, fixing a renormalization scheme (``match'') for the
effective theory. The details of the matching conditions should
of course be irrelevant, as long as they fix the renormalization
constants. To match the axial current, we choose the condition
\be
  \label{e_MatchCond}
  \Xmatch(\mMSbar)=Y_{\rm CA}(z,a/L)+{\rm O}(1/z)+{\rm O}((a/L)^2),
\ee
which means that the effective theory and full QCD are matched at the
scale $\mu=\mMSbar$.

The renormalized static axial current in this scheme is related to the
current in the lattice MS scheme by finite renormalization, i.\ e.\
\be
  \Xmatch(\mu)=\chiAstat\XLat(\mu)+{\rm O}((a/L)^2),
\ee
with
\be
  \chiAstat=1+\BAstat\gLat^2+{\rm O}(\glat^4),
\ee
where $\BAstat$ is a finite constant. Using
\bes
  X_{\rm lat}^{(1)}(\mMSbar) & = &  X_{\rm lat}^{(1)}(z/L) \nonumber\\
  & = & X_{\rm lat}^{(1)}(1/L)-\gamma_0\ln(z)X^{(0)}(a/L)+{\rm O}((a/L)^2),
\ees
this constant can be calculated from the
matching condition \eq{e_MatchCond},
\bes
  \BAstat X^{(0)}(a/L)&=&Y_{\rm CA}^{(1)}(z,a/L)-X_{\rm lat}^{(1)}(1/L)
  +\gamma_0\ln(z)X^{(0)}(a/L)\nonumber\\
  & & +{\rm O}(1/z)+{\rm O}((a/L)^2).
\ees
Here it is understood that all quantities are extrapolated to the
continuum limit first, and then the extrapolation to $1/z=0$ is performed
to determine $\BAstat$. In practical terms, we define a quantity 
$\hat{B}_{\rm A}^{\rm stat}(z)$ as the continuum limit of
\bes
  \hat{B}_{\rm A}^{\rm stat}(z,a/L) & = &
  \gamma_0\ln(z)
  +{1\over{X^{(0)}(a/L)}}\left\{
  Y_{\rm I}^{(1)}(z,a/L)\right. \nonumber\\
  & & \left.+Z_{\rm A}^{(1)}Y^{(0)}(z,a/L)
  -X_{\rm lat}^{(1)}(a/L)\right\}.
\ees

\begin{figure}
  \noindent
  \begin{center}
  \begin{minipage}[b]{.8\linewidth}
    \centering\epsfig{figure=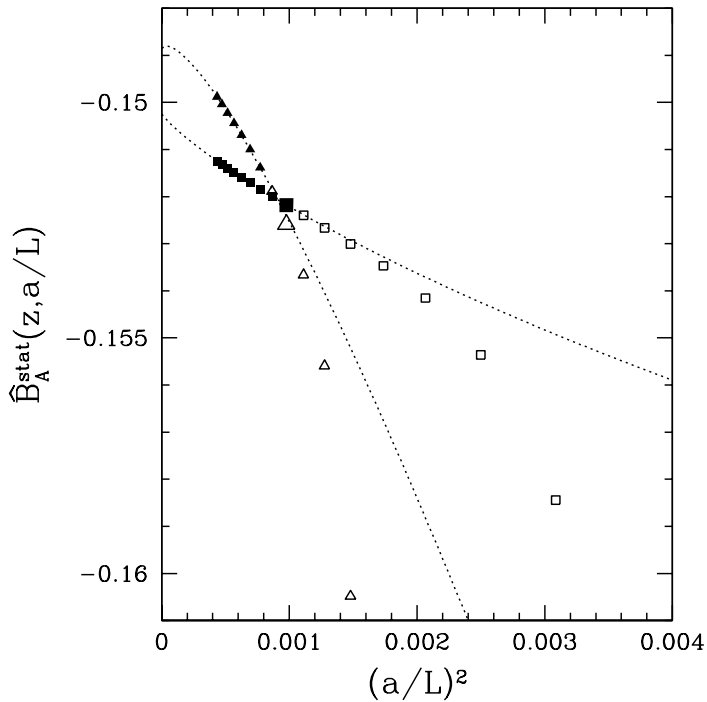,width=\linewidth}
  \end{minipage}
  \end{center}
  \caption[Continuum extrapolation of $\hat{B}_{\rm A}^{\rm stat}(z,a/L)$]
  {\footnotesize $\hat{B}_{\rm A}^{\rm stat}(z,a/L)$ 
  at $\theta=0.5$, for $z=8$ (squares) and
  $z=12$ (triangles). The bigger symbols are the data for $L/a=32$.
  \vspace{0.6cm}}
  \label{f_ContExpol}
\end{figure}

\begin{figure}
  \noindent
  \begin{center}
  \begin{minipage}[b]{.8\linewidth}
    \centering\epsfig{figure=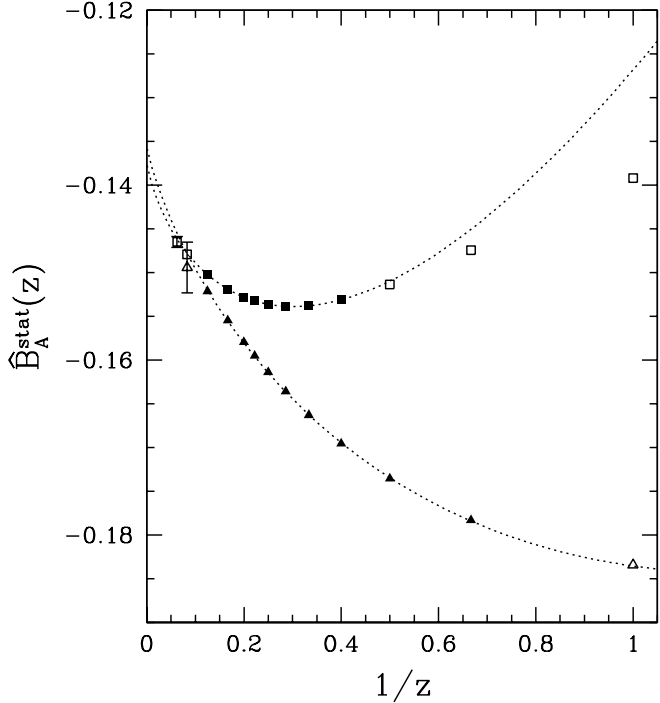,width=\linewidth}
  \end{minipage}
  \end{center}
  \caption[Calculation of $\BAstat$]{\footnotesize
    Calculation of the matching
    coefficient $\BAstat$ by
    extrapolation to $1/z=0$ for $\theta=0.0$ (triangles) and $\theta=0.5$
    (squares). Where not indicated the errors are smaller than the symbol
    sizes. \vspace{0.6cm}}
  \label{f_OneLoopFit}
\end{figure}

The continuum limit of $X^{(1)}$ and $Y^{(1)}$ is calculated separately by
fitting the numerical data with a function of the expected 
form~\cite{Luscher:1986wf}
\bes
  \label{e_CorrExp}
  h_0 & + & h_1{1\over{l^2}}+h_2{1\over {l^2}}\ln(1/l)
  \nonumber\\
  & + & h_3{1\over{l^3}}+h_4{1\over l^3}\ln(1/l)
  +{\rm O}(1/l^4),\qquad l=L/a,
\ees
using the fit procedure described in reference~\cite{Bode:1999sm}. This
reference also
explains a method to estimate the errors of the fit parameters, which we
used in our calculations. For $\theta=0.5$, we used
\eq{e_CorrExp} without the ${\rm O}((a/L)^3)$ terms for $z>5$ and included
those terms in all other fits. For some parameters, the $\hat{B}_{\rm
A}^{\rm stat}$ values resulting from our numerical data and
the fits are shown in \fig{f_ContExpol}.

It is expected that
\be
  \label{e_OneLoopFit}
  \hat{B}_{\rm A}^{\rm stat}(z)
  = \BAstat+\hat{f}_1{1\over{z}}
    +\hat{f}_2{1\over{z}}\ln({1\over{z}})+{\rm O}({1\over{z^2}}),
\ee
where ${\rm O}(1/z^2)$ means terms of the form $1/z^2$ and $\ln(z)/z^2$,
and the constants $\hat{f}_1$ and $\hat{f}_2$
can be determined from a fit to the 
numerical data. \Fig{f_OneLoopFit} shows the data for a set of values of
$z$
between $1$ and $12$ at $\theta=0.0$ and
between $1$ and $16$ at $\theta=0.5$, as well as fits of the form
(\ref{e_OneLoopFit}), where only the filled symbols have been included in the
fits. The results for the fit constants are

\begin{tabular}{|r|r|r|r|}
  \hline
  $\theta$ & $\BAstat$ & $\hat{f}_1$ & $\hat{f}_2$ \\
  \hline\hline
  $0.0$ & $-0.136(3)$ & $-0.05(2)$ & $-0.04(2)$ \\
  $0.5$ & $-0.137(1)$ & $0.011(5)$ & $-0.054(5)$ \\
  \hline
\end{tabular}

\noindent
which means that our final result for $\BAstat$ is
\be
  \BAstat=-0.137(1).
\ee
This constant is already known from~\cite{BorrPitt}, where it has been
calculated from matrix elements between quark states, using a gluon mass
as infrared regulator. Extracting the relevant terms from that calculation,
one gets
\be
  \BAstat=-0.14,
\ee
in agreement with our result. 

It is more illuminating to turn this logic around:
If we take the matching constant $\BAstat$
from matrix elements between quark states,
the correlation functions calculated from the static effective theory
are the infinite mass limits of the corresponding correlation functions in
full QCD, and the mass-dependence of the QCD correlation functions at one loop
level has the form expected from HQET. To our knowledge this is the first 
explicit demonstration of this fact for gauge invariant, infrared finite
correlation functions.

%%% Local Variables: 
%%% mode: latex
%%% TeX-master: "pap2"
%%% End: 

%% file: sect4.tex
\section{Heavy quark expansion as an asymptotic series --- finite
mass effects \label{s_FiniteMass}}

\begin{figure}
  \noindent
  \begin{center}
  \begin{minipage}[b]{.8\linewidth}
    \centering\epsfig{figure=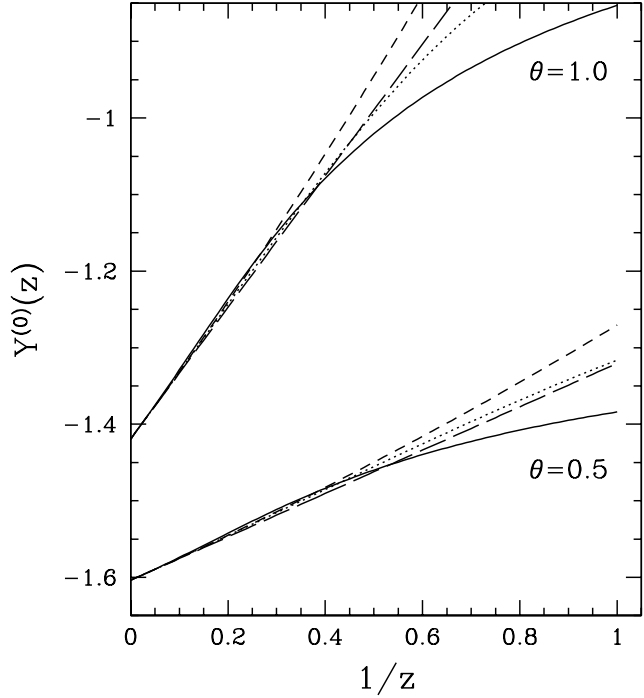,width=\linewidth}
  \end{minipage}
  \end{center}
  \caption[$Y(z)$ at tree level]{\footnotesize $Y(z)$
  at tree level (solid curve),
  for $T=L$ and two values of $\theta$, with the expansion
  up to order $1/z$ (long dashes), order $(1/z)^2$ (short dashes), and
  order $(1/z)^3$ (dotted curve). \vspace{1cm}}
  \label{f_TreeFit}
\end{figure}

As $\Xmatch$ is defined by matching to $Y_{\rm CA}$, the
static theory and QCD with relativistic quarks can be compared directly,
when these two renormalization schemes are used. 
To calculate the actual size of the finite mass effects, i.e. the difference
between the ratio $Y_{\rm CA}$ and the static limit, we define
\be
  \Delta(z)={{Y_{\rm CA}(z)-X_{\rm match}(\mMSbar)}\over{
     X_{\rm match}(\mMSbar)}},
\ee
and expand
\be
  \Delta(z)=\Delta^{(0)}(z)+\Delta^{(1)}(z)g_{\msbar}^2
     +{\rm O}(g_{\msbar}^4).
\ee

At tree level, the continuum limits of $X^{(0)}$
and $Y^{(0)}$ can be calculated ana\-lytically,
\be
  \label{e_X0Explicit}
  X^{(0)}(a/L)=-\sqrt{{{3\omega_1}\over{2R_1}}}(1+e^{-\omega_1\tau})
  +{\rm O}((a/L)^2),
\ee
\bes
  \label{e_Y0NonAnalytic}
  Y^{(0)}(z,a/L) & = & {1\over\sqrt{2\omega_2R_2}}\{(z+\omega_2)
  +(z-\omega_2)e^{-\omega_2\tau}\}X^{(0)}(a/L) \nonumber\\
  & & +{{3\sqrt{3}\theta^2}\over{2\sqrt{\omega_1\omega_2R_1R_2}}}
   (1-e^{-\omega_1\tau})(1-e^{-\omega_2\tau})+{\rm O}((a/L)^2),
\ees
where
\be
  \tau={T\over{L}},
\quad
  \omega_1=\sqrt{3}\theta,
\quad
  \omega_2=\sqrt{z^2+3\theta^2},
\ee
\be
  R_1=\omega_1(1+e^{-2\omega_1 \tau}),
\quad
  R_2=z+\omega_2-(z-\omega_2)e^{-2\omega_2 \tau}.
\ee
\Eq{e_Y0NonAnalytic} shows that the continuum limit of
$Y^{(0)}$ is non-analytic in $1/z$, which
means that an expansion in $1/z$ in the sense of a convergent Taylor series
is not possible, the expansion is at most asymptotic. However, the
non-analytic terms in \eq{e_Y0NonAnalytic} are suppressed by terms of the
form $e^{-\omega_2T}$, and in the static limit $z\rightarrow\infty$, the
ratio $Y^{(0)}$ equals the ratio $X^{(0)}$ calculated from the static
quark action \eq{e_ContLag}.
While $Y^{(0)}(z)=X^{(0)}$ for $\theta=0$, there are finite mass effects
for other values of $\theta$.
For $\theta=0.5$ and $\theta=1.0$ the resulting
curves for $Y^{(0)}$ as a function of $1/z$ are shown in \fig{f_TreeFit},
together with the expansion up to order $(1/z)^3$.
The $1/z\rightarrow 0$ limit of these plots gives the respective value of
$X^{(0)}$.
It is clearly visible
that the heavy quark expansion breaks down for $z<2$, which corresponds to
$\mMSbar<2/L$.

From \eq{e_X0Explicit} and \eq{e_Y0NonAnalytic}, we get the tree level
quantity $\Delta^{(0)}$. The result
at $T=L$ is that the finite mass effects for $z\geq 1$ are in the range of
0--14 per cent for $\theta=0.5$ and 0--40 per cent for $\theta=1.0$.
Details can be read off from \fig{f_TreeFit}.

At one-loop level, the correlation functions are evaluated numerically as
described above,
and $\Delta^{(1)}(z)$ is calculated from the numerical data
shown in ~\fig{f_OneLoopFit}.
For the set of parameters used in \fig{f_OneLoopFit}, $\Delta^{(1)}$ lies
between 0 and 5 per cent for $\theta=0.0$, and between 0 and 2 per cent for
$\theta=0.5$.

%%% Local Variables: 
%%% mode: latex
%%% TeX-master: "pap2"
%%% End: 

%% file: sect5.tex
\section{Cutoff effects \label{s_Cutoff}}

\begin{figure}[!b]
  \noindent
  \begin{center}
  \begin{minipage}[b]{.8\linewidth}
    \centering\epsfig{figure=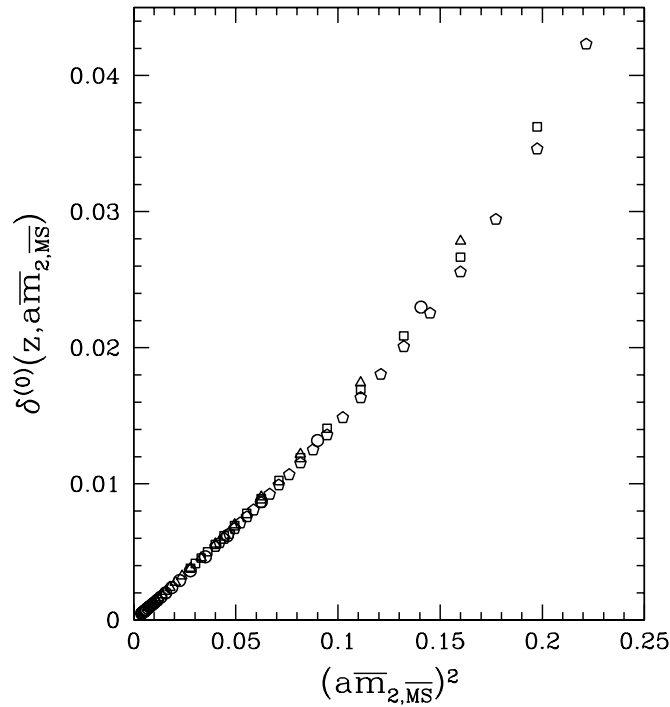,width=\linewidth}
  \end{minipage}
  \end{center}
  \caption[Discretization errors at tree level]{\footnotesize Discretization
    errors in the ratio $Y(z)$ at tree level. The circles are $z=3$ data,
    the triangles $z=4$, the squares $z=8$, and the pentagons $z=16$.}
  \label{f_DiscErrTree}
\end{figure}

\begin{figure}
  \noindent
  \begin{center}
  \begin{minipage}[b]{.8\linewidth}
    \centering\epsfig{figure=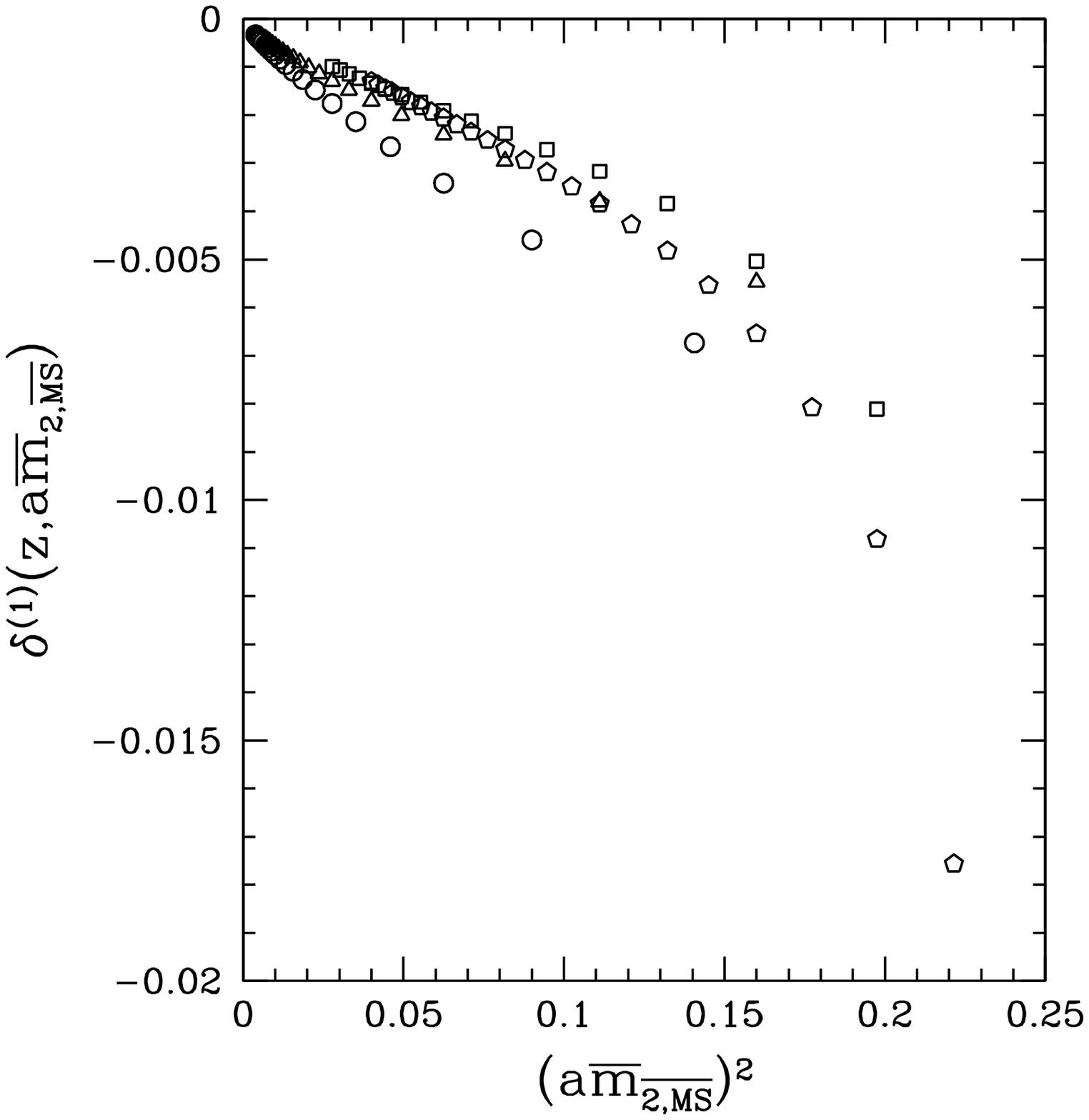,width=\linewidth}
  \end{minipage}
  \end{center}
  \caption[Discretization errors at one-lop level]{\footnotesize
    Discretization
    errors in the ratio $Y(z)$ at one-loop level. The circles are $z=3$ data,
    the triangles $z=4$, the squares $z=8$, and the pentagons $z=16$.
    \vspace{0.6cm}}
  \label{f_DiscErrOneLoop}
\end{figure}

\begin{figure}
  \noindent
  \begin{center}
  \begin{minipage}[b]{.8\linewidth}
    \centering\epsfig{figure=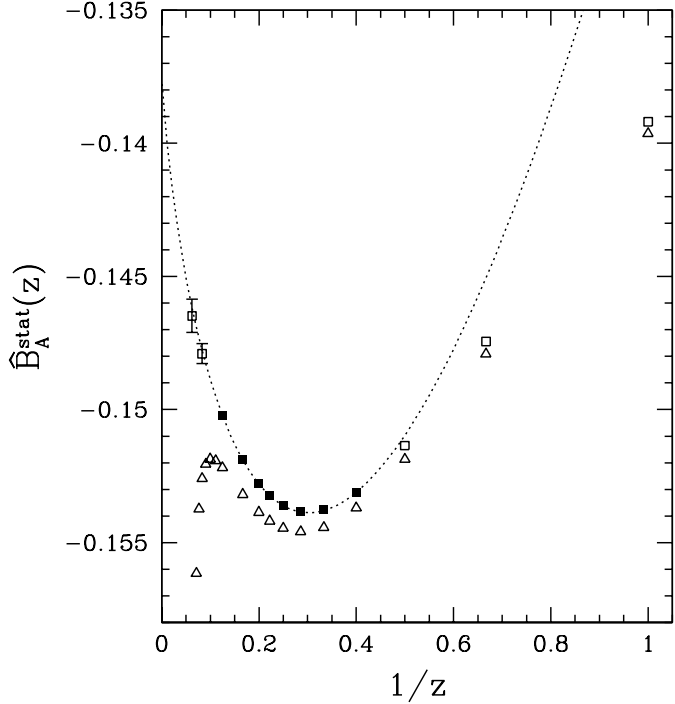,width=\linewidth}
  \end{minipage}
  \end{center}
  \caption[Discretization errors in $\hat{B}_{\rm A}^{\rm
    stat}$]{\footnotesize The matching
    coefficient $\hat{B}_{\rm A}^{\rm stat}$ 
    in the continuum limit (squares) and at $L/a=32$ (triangles)
    for $\theta=0.5$. The dotted line is the fit explained in
    \sect{s_RenMatch}. \vspace{0.6cm}}
  \label{f_L32Plot}
\end{figure}

As we use an ${\rm O}(a)$ improved action, improved boundary fields, and
improved operators in our lattice calculation, we expect discretization
errors of ${\rm O}(a^2)$. These are expected to depend on the heavy quark
mass. To shed some light on the mass dependence of the discretization
errors, we define
\be
  \delta(z,a\mMSbar)={{Y_{\rm CA}(z,a/L)-Y_{\rm CA}(z,0)}
    \over{Y_{\rm CA}(z,0)}},
\ee
and expand again,
\be
  \delta=\delta^{(0)}
    +\delta^{(1)}g_{\msbar}^2+{\rm O}(g_{\msbar}^4).
\ee
We calculate $\delta^{(0)}$ and $\delta^{(1)}$ for a range of different
values of $z$ and $a/L$, and print them as a function of $(a\mMSbar)^2$
in \fig{f_DiscErrTree} and \fig{f_DiscErrOneLoop}. The figures show that
(for our choice of parameters) 
the discretization errors are roughly proportional
to $(a\mMSbar)^2$. This is a significant problem in heavy quark
extrapolations at finite lattice spacing, as due to the mass dependence of
the discretization errors, there is a certain risk that the slope being
used in the extrapolation contains a sizeable contribution from
discretization errors. 
To illustrate this a bit further, \fig{f_L32Plot} shows the quantity
$\hat{B}_{\rm A}^{\rm stat}$ introduced in \sect{s_RenMatch}, both in the
continuum with the fit described above, and at finite lattice spacing
$L/a=32$. Here again it can be seen that the discretization errors increase
with the value of the heavy quark mass, and for $z>8$ (which means 
$a\mMSbar>1/4)$) the behaviour of the
curve at finite lattice spacing changes dramatically. That this is really
due to a cutoff effect can be seen from \fig{f_ContExpol}, where the data
points for $L/a=32$ have been marked by slightly bigger symbols.
This shows that ${\rm O}(a)$ improvement breaks down for too large values
of $a\mMSbar$, and heavy quark extrapolations should
be performed in the continuum limit rather than at finite lattice spacing.
A detailed explanation of the improvement breakdown for heavy quarks can be
found in section 4 of 
ref.~\cite{Sint:1996ch}, where this effect is discussed in the context of
the Schr\"odinger functional renormalization of the QCD gauge coupling.

A comparison between the values of
$\delta(z,a\mMSbar)$ for $a\mMSbar<1/2$
and $\Delta(z)$ for $z\geq 2$,
both at
tree and one-loop level, shows that the discretization errors and the 
finite mass effects are of comparable size, i.e. of
the order of a few per cent.

%% file: sect6.tex
\section{Conclusions \label{s_Conclusions}}

The main point of this paper is a comparison between the axial current in
QCD with finite quark
masses, and in the static approximation at one-loop order of perturbation
theory. To achieve this, we have defined gauge invariant, infrared finite,
finite-volume correlation functions
with Schr{\"o}dinger functional boundary conditions, and constructed
ratios of these correlation functions to cancel renormalization constants
except for that of the axial current. Furthermore, the divergence due to
the residual mass term in the static theory cancels in our ratios.

Using
these ratios, and assuming renormalizability of the static theory,
we have defined a renormalization scheme for the static-light
axial current by matching to the full theory, such that the renormalized
static ratio equals the heavy-quark limit of the corresponding ratio
calculated from full QCD. This renormalization scheme allows a comparison
between the full and the static theory.
Taking the matching constant $\BAstat$ from one specific matching condition,
we have verified explicitly that the heavy quark limit of an independent
correlation function is described by HQET.
For quark masses down to $1/L$, where
$L$ is the extent of the space-time box, the finite-mass effects in
our ratio are in the range of a few
per cent, both at tree level and at one-loop
order of perturbation theory.

As we performed our calculations in the lattice regularization, we were also
able to investigate discretization errors in our ratio of correlation 
functions, both at tree and one-loop level. We have shown that (for our
choice of pa\-rameters), the tree level and one-loop coefficients approach
the continuum limit at a rate roughly proportional to $(am)^2$, if
$\Oa$ improvement is applied to the action, the boundary fields, and to
composite operators. We observe a significant difference between continuum
limit data and results at finite lattice spacing, with a qualitative
difference in the behaviour of the two curves for $am>1/4$. This
demonstrates that extrapolations in the heavy quark mass should be carried
out in the continuum limit rather than at finite lattice spacings.

%% file: sect7.tex
\section*{Acknowledgements \label{s_Acknowledgement}}

We thank F.~Jegerlehner for useful comments on the manuscript.
This work is supported by the European Community's Human potential
programme under HPRN-CT-2000-00145 Hadrons/LatticeQCD.

%% file: appa.tex
\section{Feynman diagrams \label{a_Feynman}}

\begin{figure}[!h]
  \noindent
  \begin{center}
  \begin{minipage}[b]{.3\linewidth}
    \centering\epsfig{figure=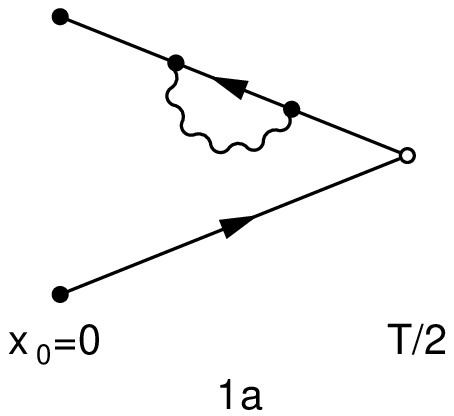,width=\linewidth}
  \end{minipage}
  \begin{minipage}[b]{.3\linewidth}
    \centering\epsfig{figure=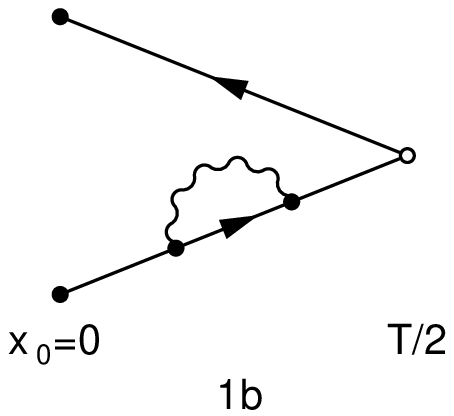,width=\linewidth}
  \end{minipage}\\
  \vspace{4mm}
  \begin{minipage}[b]{.3\linewidth}
    \centering\epsfig{figure=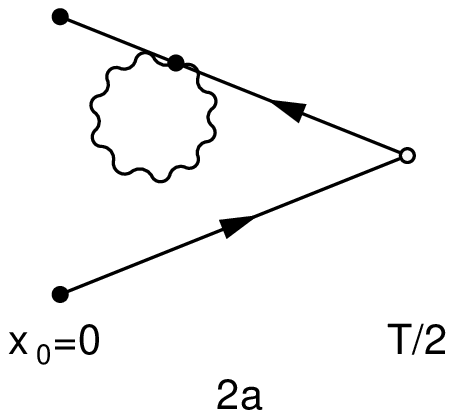,width=\linewidth}
  \end{minipage}
  \begin{minipage}[b]{.3\linewidth}
    \centering\epsfig{figure=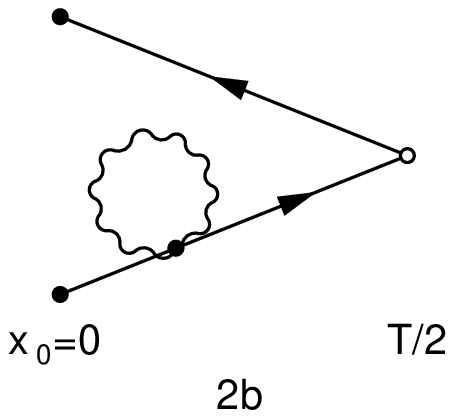,width=\linewidth}
  \end{minipage}
  \begin{minipage}[b]{.3\linewidth}
    \centering\epsfig{figure=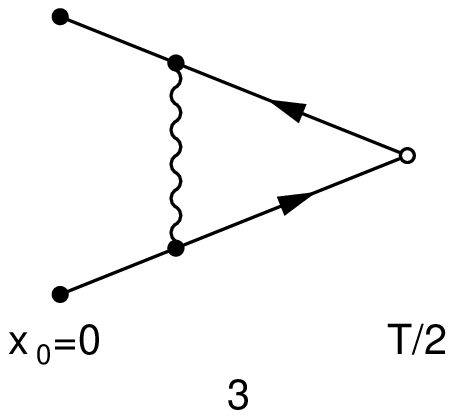,width=\linewidth}
  \end{minipage}
  \end{center}
  \caption{One loop diagrams contributing to $f_{\rm A}(T/2)$. \vspace{0.6cm}}
  \label{f_fa}
\end{figure}

\begin{figure}[!h]
  \noindent
  \begin{center}
  \begin{minipage}[b]{.3\linewidth}
    \centering\epsfig{figure=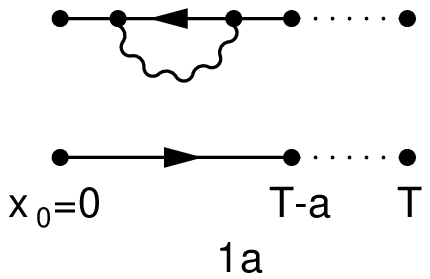,width=\linewidth}
  \end{minipage}
  \begin{minipage}[b]{.3\linewidth}
    \centering\epsfig{figure=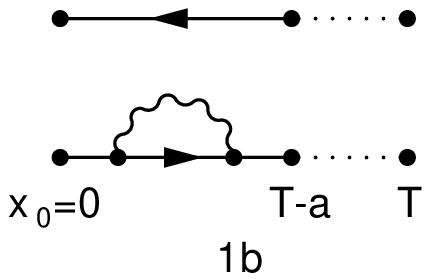,width=\linewidth}
  \end{minipage}
  \begin{minipage}[b]{.3\linewidth}
    \centering\epsfig{figure=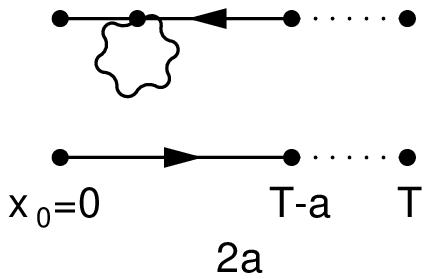,width=\linewidth}
  \end{minipage}\\
  \vspace{4mm}
  \begin{minipage}[b]{.3\linewidth}
    \centering\epsfig{figure=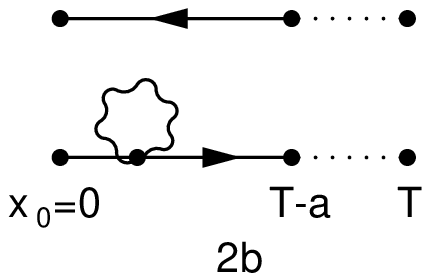,width=\linewidth}
  \end{minipage}
  \begin{minipage}[b]{.3\linewidth}
    \centering\epsfig{figure=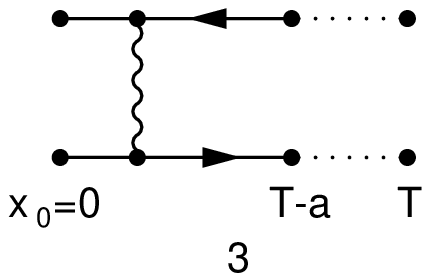,width=\linewidth}
  \end{minipage}
  \begin{minipage}[b]{.3\linewidth}
    \centering\epsfig{figure=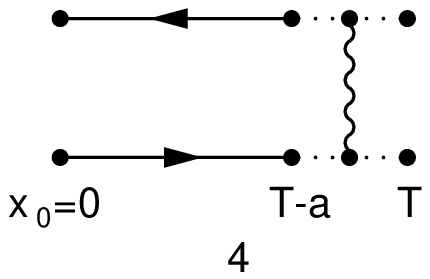,width=\linewidth}
  \end{minipage}\\
  \vspace{4mm}
  \begin{minipage}[b]{.3\linewidth}
    \centering\epsfig{figure=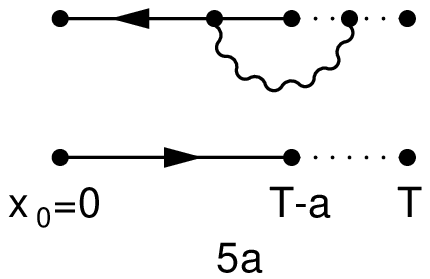,width=\linewidth}
  \end{minipage}
  \begin{minipage}[b]{.3\linewidth}
    \centering\epsfig{figure=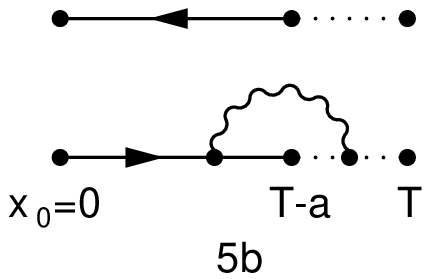,width=\linewidth}
  \end{minipage}
  \begin{minipage}[b]{.3\linewidth}
    \centering\epsfig{figure=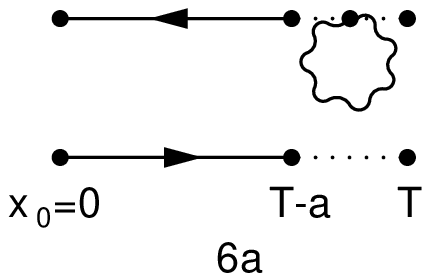,width=\linewidth}
  \end{minipage}\\
  \vspace{4mm}
  \begin{minipage}[b]{.3\linewidth}
    \centering\epsfig{figure=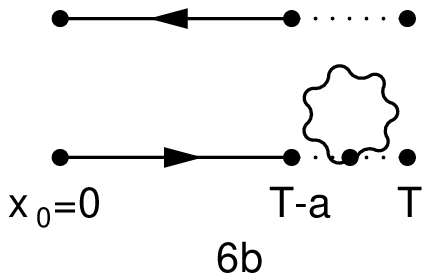,width=\linewidth}
  \end{minipage}
  \begin{minipage}[b]{.3\linewidth}
    \centering\epsfig{figure=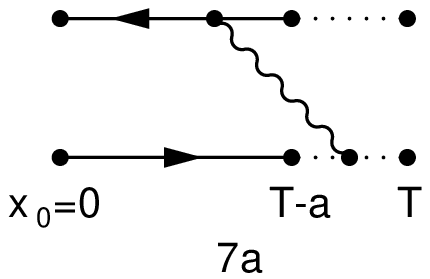,width=\linewidth}
  \end{minipage}
  \begin{minipage}[b]{.3\linewidth}
    \centering\epsfig{figure=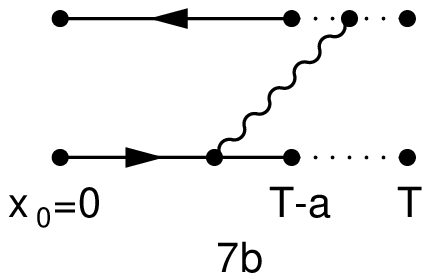,width=\linewidth}
  \end{minipage}
  \end{center}
  \caption{One loop diagrams contributing to $f_1$. The
    dotted lines denote the link from $T-a$ to T. \vspace{0.6cm}}
  \label{f_f1}
\end{figure}

The expansion of the correlation functions $f_{\rm A}$ and $f_1$ is
described in detail in ref.~\cite{impr:pap2}. Following that method, we
get the one-loop Feynman diagrams shown in \fig{f_fa} and \fig{f_f1}.
\pagebreak

%% file: pap2.bbl
\begin{thebibliography}{10}

\bibitem{Neubert:1994mb}
M. Neubert,
\newblock Phys. Rept. 245 (1994) 259, hep-ph/9306320,
\newblock and references therein.

\bibitem{Kurth:2000ki}
ALPHA, M. Kurth and R. Sommer,
\newblock Nucl. Phys. B597 (2001) 488, hep-lat/0007002.

\bibitem{impr:lett}
K. Jansen et~al.,
\newblock Phys. Lett. B372 (1996) 275, hep-lat/9512009.

\bibitem{impr:Sym1}
K. Symanzik,
\newblock Nucl. Phys. B226 (1983) 187.

\bibitem{stat:eichhill1}
E. Eichten and B. Hill,
\newblock Phys. Lett. B234 (1990) 511.

\bibitem{SF:LNWW}
M. {L\"uscher}, R. Narayanan, P. Weisz and U. Wolff,
\newblock Nucl. Phys. B384 (1992) 168, hep-lat/9207009.

\bibitem{SF:stefan1}
S. Sint,
\newblock Nucl. Phys. B421 (1994) 135, hep-lat/9312079.

\bibitem{impr:pap4}
M. {L\"uscher}, S. Sint, R. Sommer and H. Wittig,
\newblock Nucl. Phys. B491 (1997) 344, hep-lat/9611015.

\bibitem{impr:pap5}
S. Sint and P. Weisz,
\newblock Nucl. Phys. B502 (1997) 251, hep-lat/9704001.

\bibitem{MORNINGSTAR1}
C. Morningstar and J. Shigemitsu,
\newblock Phys. Rev. D57 (1998), hep-lat/9712015.

\bibitem{Ishikawa}
K.I. Ishikawa, T. Onogi and N. Yamada,
\newblock Nucl. Phys. Proc. Suppl. 83 (2000) 301, hep-lat/9909159.

\bibitem{impr:pap2}
M. {L\"uscher} and P. Weisz,
\newblock Nucl. Phys. B479 (1996) 429, hep-lat/9606016.

\bibitem{Reisz:1989kk}
T. Reisz,
\newblock Nucl. Phys. B318 (1989) 417.

\bibitem{Shifman:1987sm}
M.A. Shifman and M.B. Voloshin,
\newblock Sov. J. Nucl. Phys. 45 (1987) 292.

\bibitem{Politzer:1988wp}
H.D. Politzer and M.B. Wise,
\newblock Phys. Lett. B206 (1988) 681.

\bibitem{pert:gabrielli}
E. Gabrielli, G. Martinelli, C. Pittori, G. Heatlie and C.T. Sachrajda,
\newblock Nucl. Phys. B362 (1991) 475.

\bibitem{Boch}
M. Bochicchio, L. Maiani, G. Martinelli, G.C. Rossi and M. Testa,
\newblock Nucl. Phys. B262 (1985) 331.

\bibitem{Mannel:1992mc}
T. Mannel, W. Roberts and Z. Ryzak,
\newblock Nucl. Phys. B368 (1992) 204.

\bibitem{Luscher:1986wf}
M. {L\"uscher} and P. Weisz,
\newblock Nucl. Phys. B266 (1986) 309.

\bibitem{Bode:1999sm}
ALPHA, A. Bode, P. Weisz and U. Wolff,
\newblock Nucl. Phys. B576 (2000) 517, hep-lat/9911018.

\bibitem{BorrPitt}
A. Borrelli and C. Pittori,
\newblock Nucl. Phys. B385 (1992) 502.

\bibitem{Sint:1996ch}
S. Sint and R. Sommer,
\newblock Nucl. Phys. B465 (1996) 71, hep-lat/9508012.

\end{thebibliography}
